\documentstyle[12pt,epsfig,a4,rotating]{article}
\def\Journal#1#2#3#4{{#1} {\bf #2}, #3 (#4)}

\def\NIMA{{\em Nucl. Instrum. Methods} A}

\def\NPBP{{\em Nucl. Phys.} B (Proc. Suppl.)}
\def\PLB{{\em Phys. Lett.}  B}
\def\PRL{\em Phys. Rev. Lett.}
\def\PRD{{\em Phys. Rev.} D}

\def\RPP{\em Rep. Prog. Phys.}

\def\ra{\rightarrow}
\def\be{\begin{equation}}  
\def\ee{\end{equation}}
\def\bea{\begin{eqnarray}}
\def\eea{\end{eqnarray}}
\newcommand{\lsnd}{LSND } 
\newcommand{\nue}{neutrino } 
\newcommand{\nues}{neutrinos }
\newcommand{\hdm}{hot dark matter }

\newcommand{\noszp}{neutrino oscillations. }
\newcommand{\osz}{oscillation }
\newcommand{\oszsp}{oscillations. }
\newcommand{\oszs}{oscillations }

\newcommand{\sk}{Superkamiokande }

\newcommand{\ssm}{see-saw-mechanism }
\newcommand{\adn}{almost degenerated neutrinos }
\newcommand{\delm}{\mbox{$\Delta m^2$} }
\newcommand{\me}{\mbox{$m_{\nu_e}$} }
\newcommand{\mmu}{\mbox{$m_{\nu_\mu}$} }
\newcommand{\mtau}{\mbox{$m_{\nu_\tau}$} }
\newcommand{\nel}{\mbox{$\nu_e$} }
\newcommand{\bnel}{\mbox{$ \bar \nu_e$} }
\newcommand{\bnmu}{\mbox{$ \bar \nu_{\mu}$} }
\newcommand{\nmu}{\mbox{$\nu_\mu$} }
\newcommand{\ntau}{\mbox{$\nu_\tau$} }
\newcommand{\sint}{\mbox{$sin^2 2\theta$} }
\newcommand{\lbls}{long baseline experiments }
\newcommand{\lbl}{long baseline experiment }
\newcommand{\sbl}{short baseline experiment }
\renewcommand{\baselinestretch}{1.5}
\begin{document}
{\bf STATUS OF TERRESTRIAL NEUTRINO OSCILLATION SEARCHES\footnote{to
appear in Proc. 4th Int. Solar Neutrino Conference, Heidelberg, April
1997}} \\
\\
\begin{center}
{\large K. Zuber}
\\ 
{\it Lehrstuhl f\"ur Experimentelle Physik IV, Universit\"at Dortmund,\\
44287
Dortmund, Germany}
\end{center}
\vspace{1cm}
The present status of \nue \osz searches with reactors and accelerators is reviewed.
An outlook, especially
on future long baseline \nue \osz projects, is given.
\setlength{\baselineskip}{1.5em}
\section{Introduction}
For several reasons, massive \nues are important in modern particle physics.
The theoretical extensions in
form
of grand unified theories of the very successful standard model 
may be checked by their prediction of \nue masses. 
Moreover massive \nues would open up a variety of new phenomena which could
be investigated by experiments. Also many astrophysical and cosmological implications would emerge from it.
For example, \nue \oszs are the preferred particle physics solution of the solar \nue
problem. Neutrinos in the mass region of a few eV serve as good candidates for \hdm . Experimental
evidence for effects due to \nue masses are seen in solar \nues , atmospheric \nues and the 
\lsnd experiment (see sec. 5.1).\\  
On the theoretical side, two groups of models for \nue masses emerged over the past years to
explain the
present observations of \nue experiments.
The first one is the ''classical'' quadratic \ssm, resulting in a strong scaling
behaviour of the
neutrino masses like
\be
\me : \mmu : \mtau \propto m_u^2 : m_c^2 : m_t^2
\ee   
where $m_u,m_c$ and $m_t$ are the corresponding quark masses.
Recently another type of \ssm has been coming up
resulting in more or less \adn . 
At present, on the experimental side, the direct limits for \nue masses are \cite{nmass}:
\begin{center}
\begin{tabular}{rlr}
\me $<$ &15 eV \qquad &\mbox{SN 1987a}\\
\mmu $<$ & 170 keV  \qquad &\mbox{$\pi$-decay}\\
\mtau $<$ & 18.2 MeV \qquad &\mbox{$\tau$-decay}
\end{tabular}   
\end{center}
\medskip
The current results for $\me$ taken from tritium beta decay experiments are
more stringent and
give a limit of 3.5 eV but face the problem of negative $m^2$-values.
Another bound valid only for
Majorana neutrinos results from double beta decay and is given by \cite{heimo}
\be
\langle \me \rangle = \mid \sum_i U_{ei}^2 m_i \mid < 0.5 eV
\ee
It should also be mentioned that the recently started E872 experiment at Fermilab 
will make it possible for
the first time to directly proof the existence of \ntau via weak charged current
reactions.
From the above mentioned experimental bounds on the \nue masses it becomes obvious, that a direct
 kinematic test of \mmu
and
\mtau and perhaps \me in the eV or even sub-eV-region is impossible, the only
possibility to explore this region is via \noszp
\section{Neutrino oscillations}
Similar to the quark sector, the mass eigenstates need
not to be the
same as the flavour eigenstates for \nues as well offering the possibility of oscillations.
In the simplified case of two flavours the mixing can be described by
\be
{\nu_e \choose \nu_{\mu}} = \left( \begin{array}{cc}
 cos \theta & sin \theta \\
- sin \theta & cos \theta
\end{array} \right)
{\nu_1 \choose \nu_2}
\ee
While $\sint$ describes the amplitude of the
oscillation, $\delm = m_2^2 -m_1^2$ determines the oscillation length,
characterising a full
cycle
of \osz between two flavours.
In practical units the oscillation length $L$ is given by
\be
L = \frac{4\pi E \hbar}{\Delta m^2 c^3} =
2.48 (\frac{E}{MeV})(\frac{eV^2}{\Delta m^2}) \quad m
\ee
As can be seen, \oszs do not allow an absolute mass measurement. Furthermore to
allow oscillations, the \nues must not be exactly degenerated.
Solar \nues are in a unique position to explore the small \delm - region  because of their
relative low
energy and the large distance of the sun. 
From first principles there is no preferred region
in the $\delm - \sint$
parameter space and therefore the whole has to be investigated experimentally.
\section{Reactor experiments}
On earth two artificial \nue sources exist in form of nuclear power reactors and accelerators.
\begin{figure}[hhh]
\begin{center}
\begin{tabular}{cc}
\epsfig{file=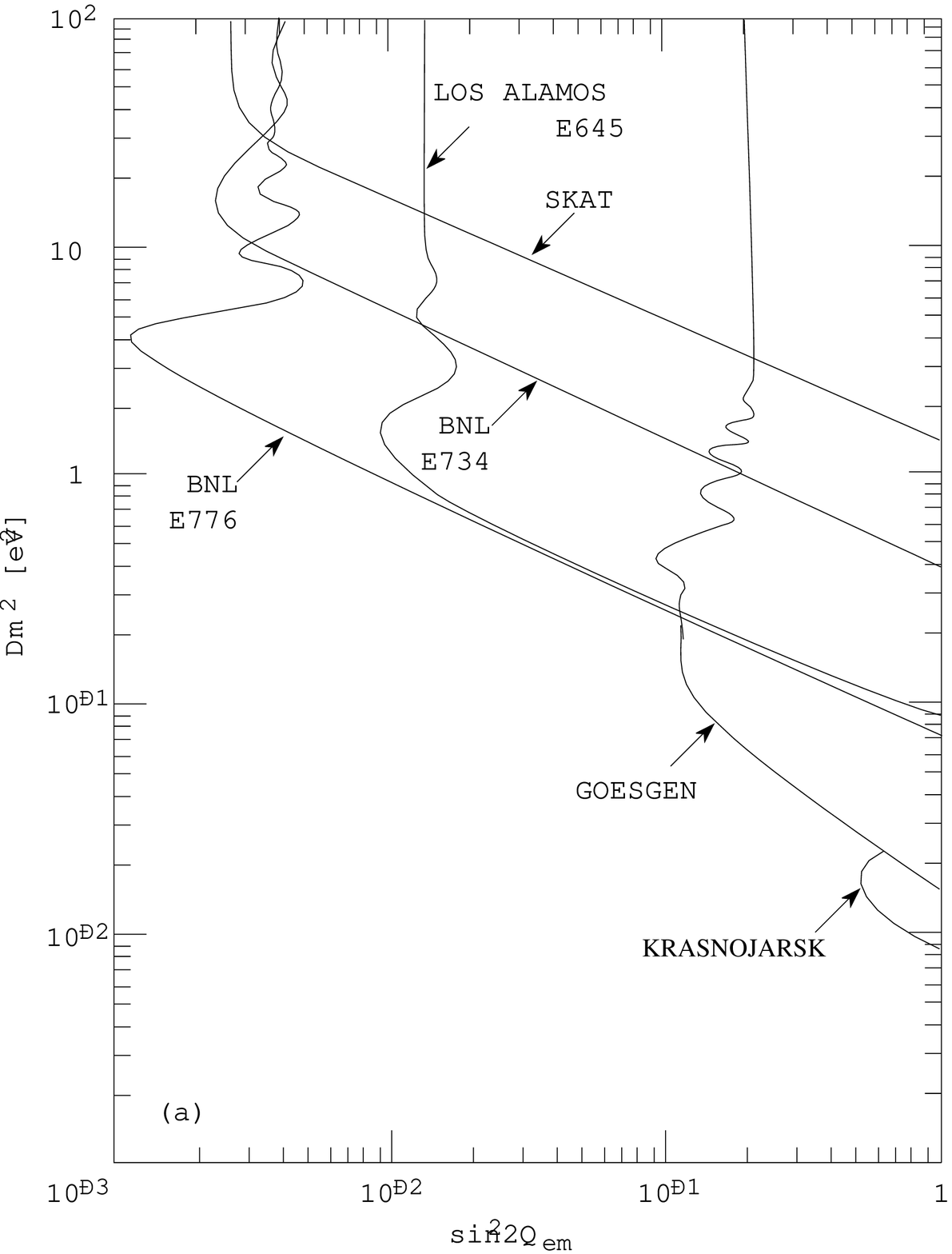,width=7.5cm,height=9cm} &
\epsfig{file=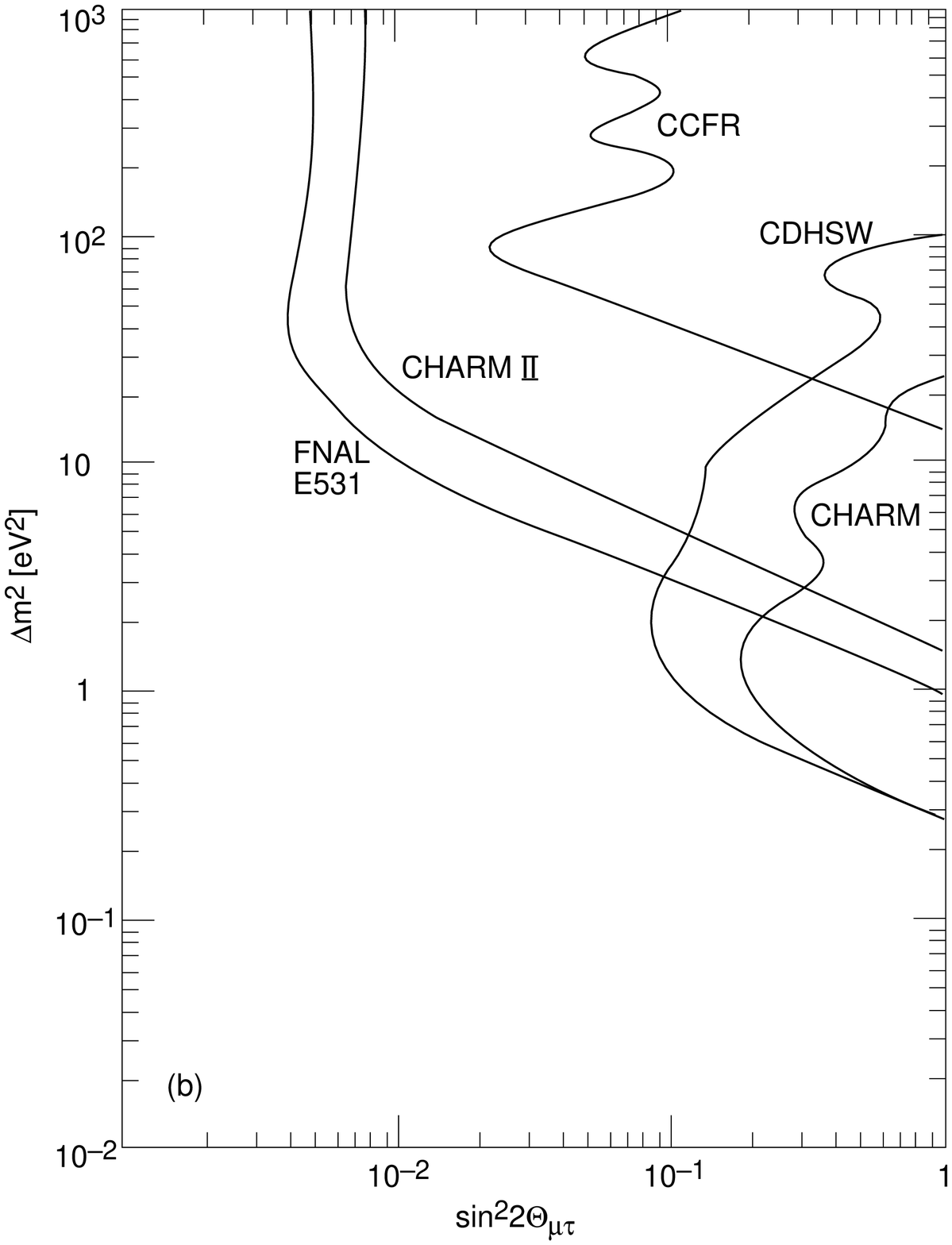,width=7.5cm,height=9cm}
\end{tabular}
\end{center}
\renewcommand{\baselinestretch}{1.}
\caption{\label{excl} \it Exclusion plots  on \nue oscillation parameters from different
reactor and
accelerator experiments. Shown are \nel - \nmu (left) and \nmu - \ntau  (right)
\oszsp The regions to the right side are excluded (from [5]).}
\end{figure}
\subsection{Principles}
Reactors are a source of MeV \bnel due to the fission of nuclear fuel. The main isotopes involved are
$^{235}$U,$^{238}$U,$^{239}$Pu and $^{241}$Pu. The neutrino rate per fission has been measured \cite{isoto}
for all isotopes except $^{238}$U and is in good agreement with
theoretical calculations \cite{kmetz}. 
Experiments typically 
try to measure the positron spectrum which can be deduced from the \bnel spectrum
and either compare it directly to the theoretical predictions
or measure it at several distances from the reactor and search for spectral changes. Both
types of experiments were done in the past. The detection reaction
is 
\be
\label{gl1}
\bnel + p \ra e^+ + n
\ee
with an energy threshold of 1.804 MeV. Reactor experiments are disappearance
experiments looking for
\bnel $\ra \bar {\nu}_X$.
\subsection{Status}
The detection reaction (\ref{gl1}) is always the same, resulting in different strategies for the detection of the
positron and the neutron. Normally coincidence techniques are used between the annihilation photons and the neutrons 
which diffuse and
thermalise within 10-100 $\mu$s.
The main background are cosmic ray muons producing neutrons 
in the surrounding of the detector.  
Several reactor experiments have been done in the past (see Tab.1). All these experiments
had a fiducial mass of less than 0.5 t and the distance to the reactor was never
more than 250 m.
\begin{center}
\begin{tabular}{|c|c|c|}
\hline
reactor & thermal power [MW] & distance [m]\\
\hline
ILL-Grenoble (F) & 57 & 8.75 \\
Bugey (F) & 2800 & 13.6, 18.3 \\
Rovno (USSR) & 1400 & 18.0,25.0 \\
Savannah River (USA) & 2300 & 18.5,23.8\\
G\"osgen (CH) & 2800 & 37.9, 45.9, 64.7 \\
Krasnojarsk (Russia) & ? & 57.0, 57.6, 231.4 \\
Bugey III (F) & 2800 & 15.0, 40.0, 95.0 \\
\hline
\end{tabular}
\medskip\\
{\it Tab.1: List of done reactor experiments. Given are the thermal power of the
reactors and the distance of
the experiments with respect to the reactor.}
\end{center} 
The achievements of the most stringent reactor experiments are shown as an exclusion plot in
Fig. \ref{excl}.
\begin{figure}[hhh]
\begin{center}
\epsfig{file=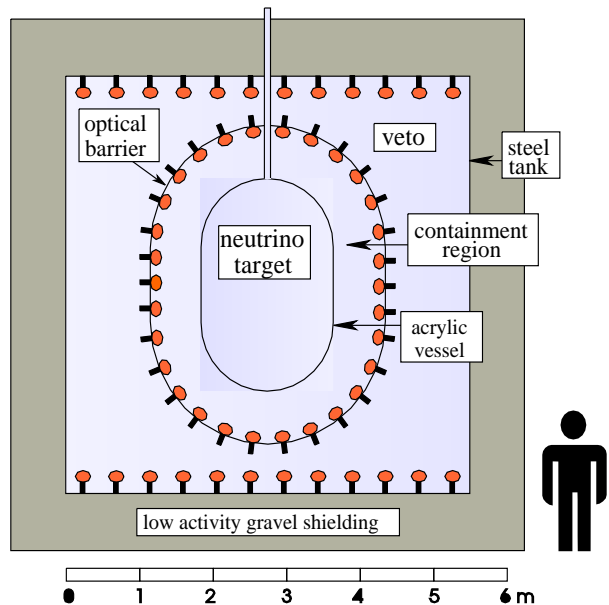,width=8cm,height=8cm}
\end{center}
\renewcommand{\baselinestretch}{1.}
\caption{\label{detcho} \it Principal design of the CHOOZ reactor experiment in France. Two
4.2 GW reactors in a distance of 
1030 m are used to produce an antineutrino flux at the experiment of about 10$^{10}$cm$^{-2}s^{-1}$.}
\end{figure}
\section{Future}
Two new reactor experiments are coming up and will be in operation in 1997. The first one is the
CHOOZ-experiment in France \cite{chooz} (Fig. \ref{detcho}).
This detector has some advantages with respect to previous experiments. First of all the detector is located underground with
a shielding of 300 mwe, reducing the background due to cosmics by a factor of 300. Moreover, the detector is about 1030 m away
from the reactor (more than a factor 4 in comparison to previous experiments)
enlarging the sensitivity to smaller \delm . In addition
the main target has about 4.8 t and is therefore much larger than those used before.
The main target consists of a specially developed
Gd-loaded scintillator. This inner detector is surrounded by an additional
detector containing 17 t of scintillator
without Gd and 90 t of scintillator as an outer veto. The signal in the inner detector will be the detection of the
annihilation photons in coincidence with n-capture on Gd, the latter producing gammas with a total sum of up to 8 MeV. 
The light from the inner region is detected by 160 photomultipliers.
Calibration runs were done in September and October 1996. The measured background is lower than the expected 2.5 
counts per day (cpd) so the expected signal of 31 cpd should be clearly visible.
The two 4.2 GW reactors will be on full power in July 1997, but data taking has
already
started.\\
\begin{figure}[hhh]
\begin{center}
\begin{tabular}{cc}
\epsfig{file=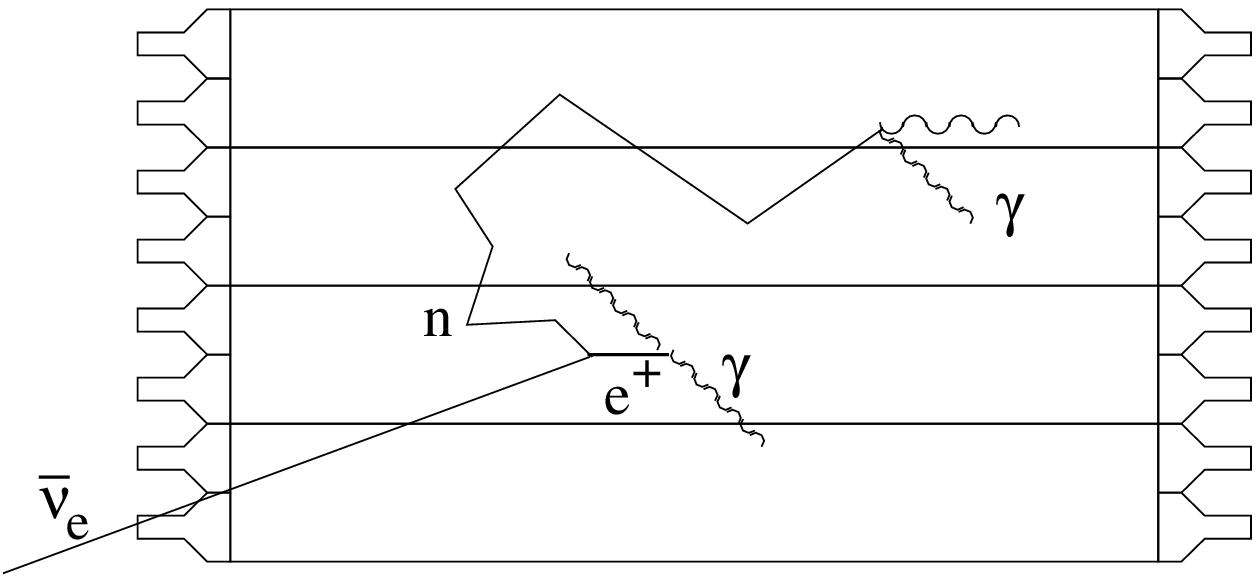,width=8cm,height=6cm} &
\epsfig{file=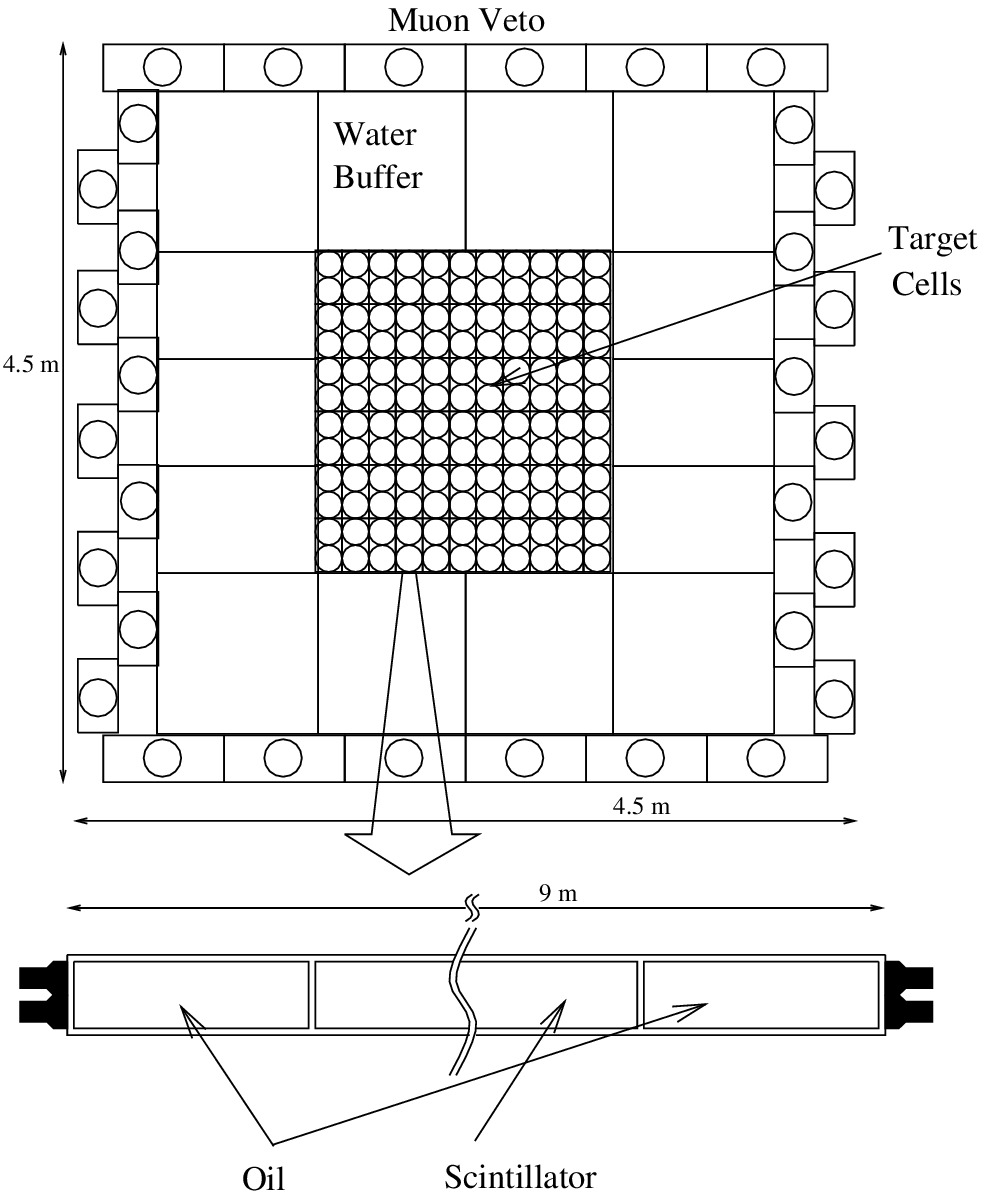,width=8cm,height=8cm}
\end{tabular}
\end{center}
\renewcommand{\baselinestretch}{1.}
\caption{\label{palov} \it Coincidence of three cells as a signal for \bnel interactions in the Palo Verde experiment (left). The principal outline of the Palo Verde detector (right). The scintillator cells
with a length of 9 m are arranged in a 6$\times$11 array, which is surrounded by a water buffer and a muon veto. For background 
reasons only the inner 7.3 m are loaded with Gd. (from [6]).}
\end{figure}
The second experiment is the Palo Verde (former San Onofre) experiment \cite{pave} near
Phoenix, AZ (USA). It will consist of 12 t liquid scintillator
also loaded with Gd. The scintillator is filled in 66 modules arranged in an 11$\times$6 array (Fig. \ref{palov}).
As a signal serves the coincidence of three modules.
The experiment will be located under a shielding of 46 mwe in a distance
of about 750 (820) m to the reactors. The experiment will be online by
late summer 1997 with an expected signal of 51 cpd.\\
Both experiments should recognize a clear signal in case the atmospheric \nue problem 
is due to \nel - \nmu oscillations.
\newline
As a project for the very future plans for a 1000 t detector (Perry) are also discussed. A further large scale
reactor experiment using the former Kamioka detector in a distance of 160 km to a reactor is approved by
the Japanese Government.
\section{Accelerators}
The second source of terrestrial \nues are high energy accelerators.
Accelerators typically produce \nue beams by shooting a proton beam on a fixed target.
The produced secondary pions and kaons decay and create a
\nue beam dominantly consisting of \nmu. 
The detection mechanism is via charged weak currents
\be
\nu_i N \ra i + X  \quad i= e, \mu, \tau
\ee
where N is a nucleon and X the hadronic final state.
Depending on the intended goal, the search for
\oszs therefore requires a detector
which is capable of detecting electrons, muons and $\tau$ - leptons in the final
state.
Accelerator experiments are typical appearance experiments working in
the channels \nmu - $\nu_X$ and \nel - $\nu_X$.
\begin{figure}[hhh]
\begin{center}  
\epsfig{file=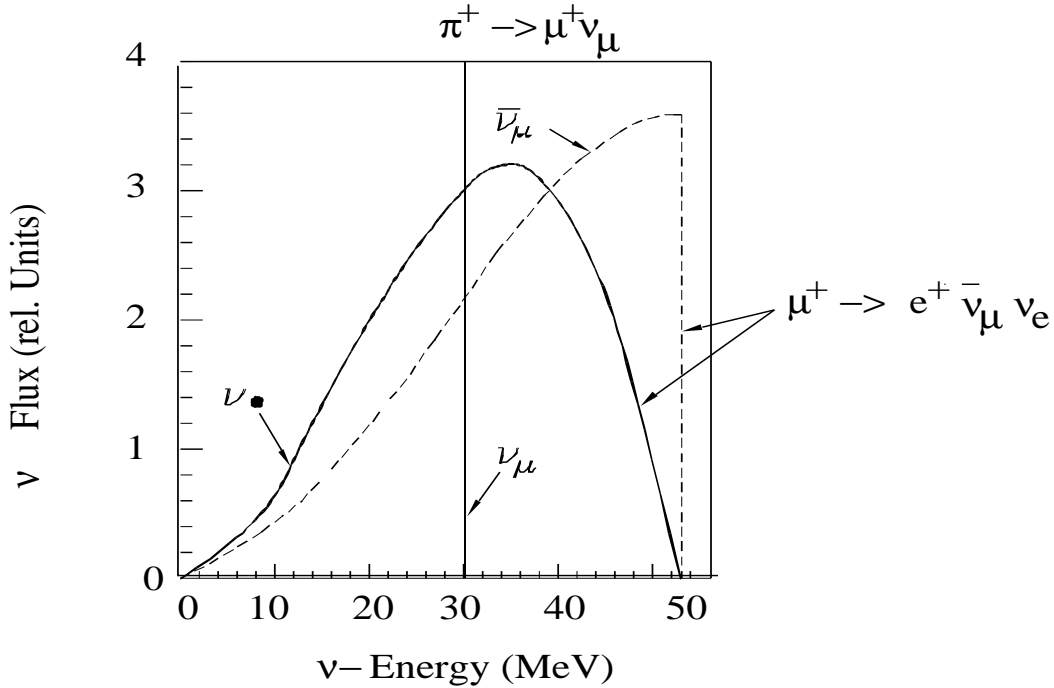,width=14cm,height=9cm}
\end{center}
\renewcommand{\baselinestretch}{1.}
\caption{\label{pispek} \it Neutrino spectrum resulting from pion and muon decays. Beside a monoenergetic
line at 29.8 MeV we can expect a continous spectrum up to 55.8 MeV. The contamination of the beam with
\bnel is normally very small.}
\end{figure}
\subsection{Accelerators at medium energy}
\begin{figure}[hhh]
\begin{center}
\epsfig{file=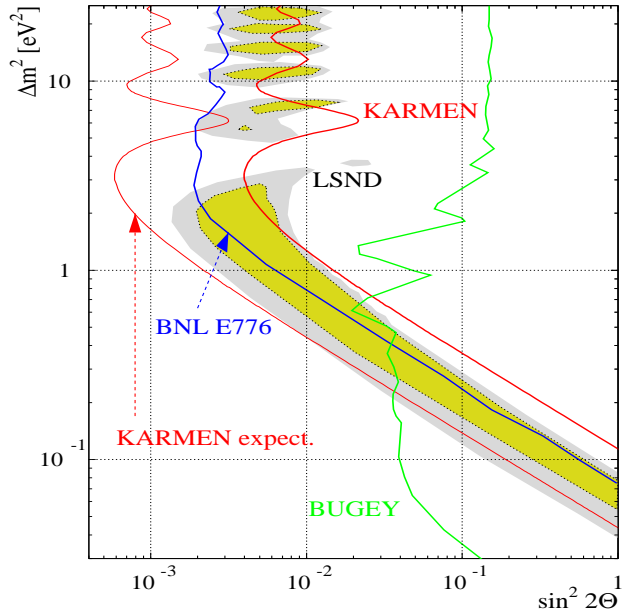,width=9cm,height=9cm}
\end{center}
\renewcommand{\baselinestretch}{1.}
\caption{\label{lsndev} \it Region of evidence for \bnmu - \bnel \oszs of \lsnd together with already excluded
parts from KARMEN, E776 and the Bugey reactor experiment. Also shown is the possibility of KARMEN after running
three more years with the new upgrade.}
\end{figure}

At present there are two experiments running with \nues at medium energies
($E_\nu \approx $ 30 - 50 MeV) namely KARMEN and \lsnd. Both experiments use
800 MeV proton beams on a beam dump to produce pions. The expected \nue
spectrum from pion and $\mu$-decay is shown in Fig. \ref{pispek}. The beam contamination 
of \bnel is in the order of $10^{-4}$. \\
The KARMEN experiment \cite{karmen} at the neutron spallation source ISIS at Rutherford
Appleton Laboratory is using 
56 t of a segmented liquid scintillator. The main
advantage of this experiment is the known time structure of the two proton pulses hitting the beam dump 
(two pulses of 100 ns with a separation of 330 ns and a repetition rate of 50 Hz).
Because of the pulsed beam positrons are expected within 0.5-10.5
$\mu$s after beam on target.
The signature for detection is a delayed coincidence of a positron in the 10 - 50 MeV region together with
$\gamma$-emission from either p(n,$\gamma$)D or Gd(n,$\gamma$)Gd reactions. The first results in 2.2 MeV photons
while the latter allow gammas up to 8 MeV. 
The limits reached so far are shown in Fig. \ref{lsndev}. 
To improve the sensitivity for \nue oscillation searches by reducing the 
neutron background a new
veto shield against atmospheric muons
was constructed which has been in operation since Feb. 1997 and is surrounding the whole detector. 
The region which can be excluded in 2-3 years of running in the upgraded version is also shown in Fig. \ref{lsndev}.
\\
The \lsnd experiment \cite{lsnddet} at LAMPF is a 167 t mineral oil based liquid
scintillation detector using
scintillation light and Cerenkov light for detection. It consists of an approximately cylindrical tank
8.3 m long and 5.7 m in diameter. The experiment is about 30 m away from a copper beam stop under an
angle of 12$^o$ with respect to the proton beam. For the \osz search in the channel 
\bnmu -\bnel a signature of a positron within the energy range 36 $<E_e<$ 60 MeV together with a
time and spatial correlated  2.2 MeV photon from p(n,$\gamma$)D is required. The analysis \cite{lsnd} ends up in evidence for
\oszs in the region shown in Fig. \ref{lsndev}.
Recently \lsnd  published their \nel - \nmu analysis for pion decays in flight \cite{lsnd1} which is in agreement with the former
evidence from pion decay at rest.
Also \lsnd continues with data aquisition.
\newline
An increase in sensitivity in the \nmu - \nel \osz channel can be reached in the future
if there is a possibility for \nue physics at the European Spallation Source (ESS) which is in the planning phase or
the National Spallation Neutron Source (NSNS) at Oak Ridge which might have a 1 GeV proton beam in 2004.
The Fermilab
8 GeV proton booster offers the chance for a \nue experiment (BooNE) as well which could start data taking in
2001.
\subsection{Accelerators at high energy}
\begin{figure}[hhh]
\begin{center} 
\epsfig{file=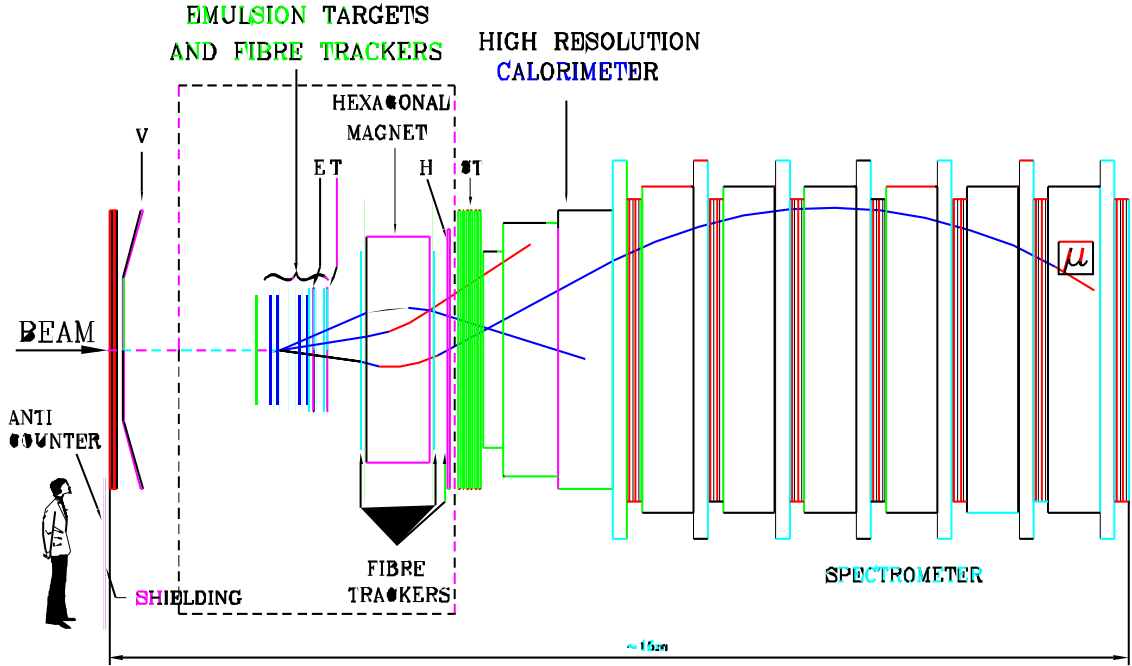,width=15cm,height=10cm}
\end{center}
\renewcommand{\baselinestretch}{1.}
\caption{\label{chorus} \it Side view of the CHORUS detector at CERN.}
\end{figure}

\begin{figure}[hhh]
\begin{center} 
\mbox{\begin{turn}{270}
\epsfig{file=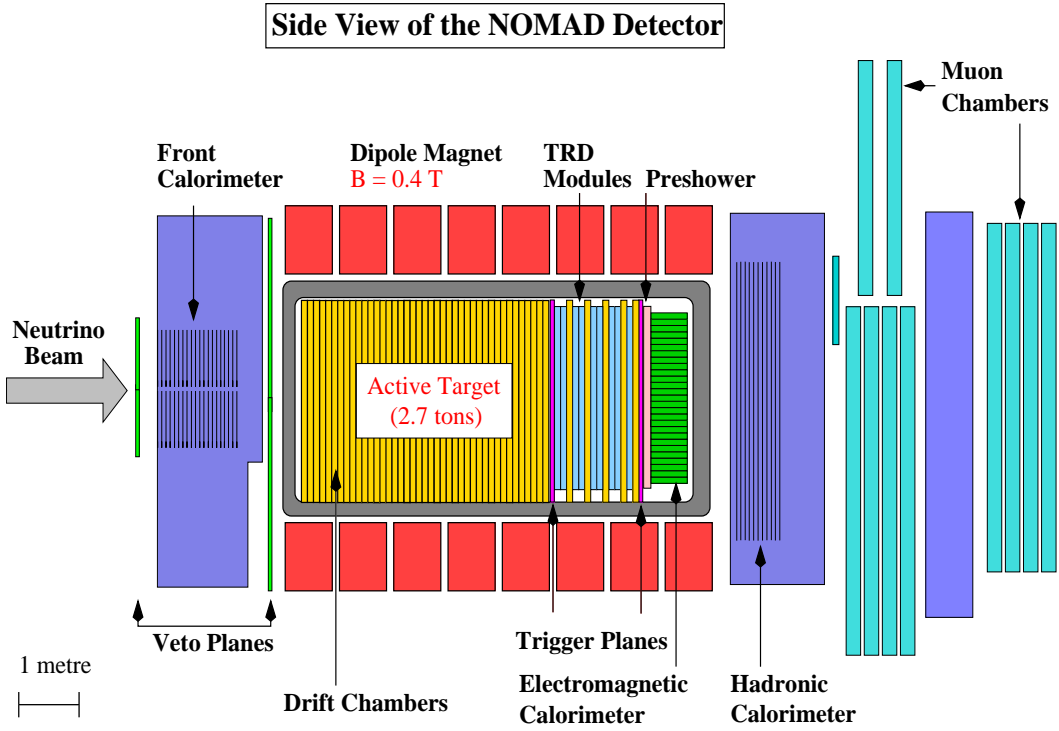,width=10cm,height=14cm}
\end{turn}}
\end{center}
\renewcommand{\baselinestretch}{1.}
\caption{\label{nomad} \it Side view of the NOMAD detector at CERN.}
\end{figure}

\begin{figure}[hhh]
\begin{center} 
\epsfig{file=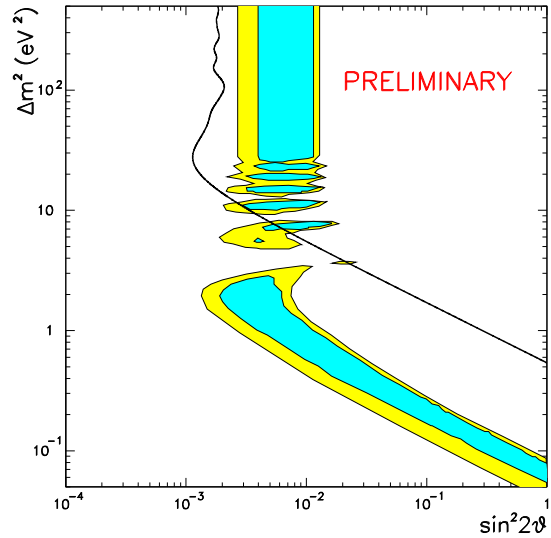,width=8cm,height=8cm}
\end{center}
\renewcommand{\baselinestretch}{1.}
\caption{\label{nomadel} \it Preliminary NOMAD exclusion plot for \nel - \nmu \osz
in comparison with \lsnd results.
The high \delm region is excluded.}
\end{figure}

\begin{figure}[hhh]
\begin{center} 
%\mbox{\begin{turn}{270}
\begin{tabular}{cc}
\mbox{\begin{turn}{270}
\epsfig{file=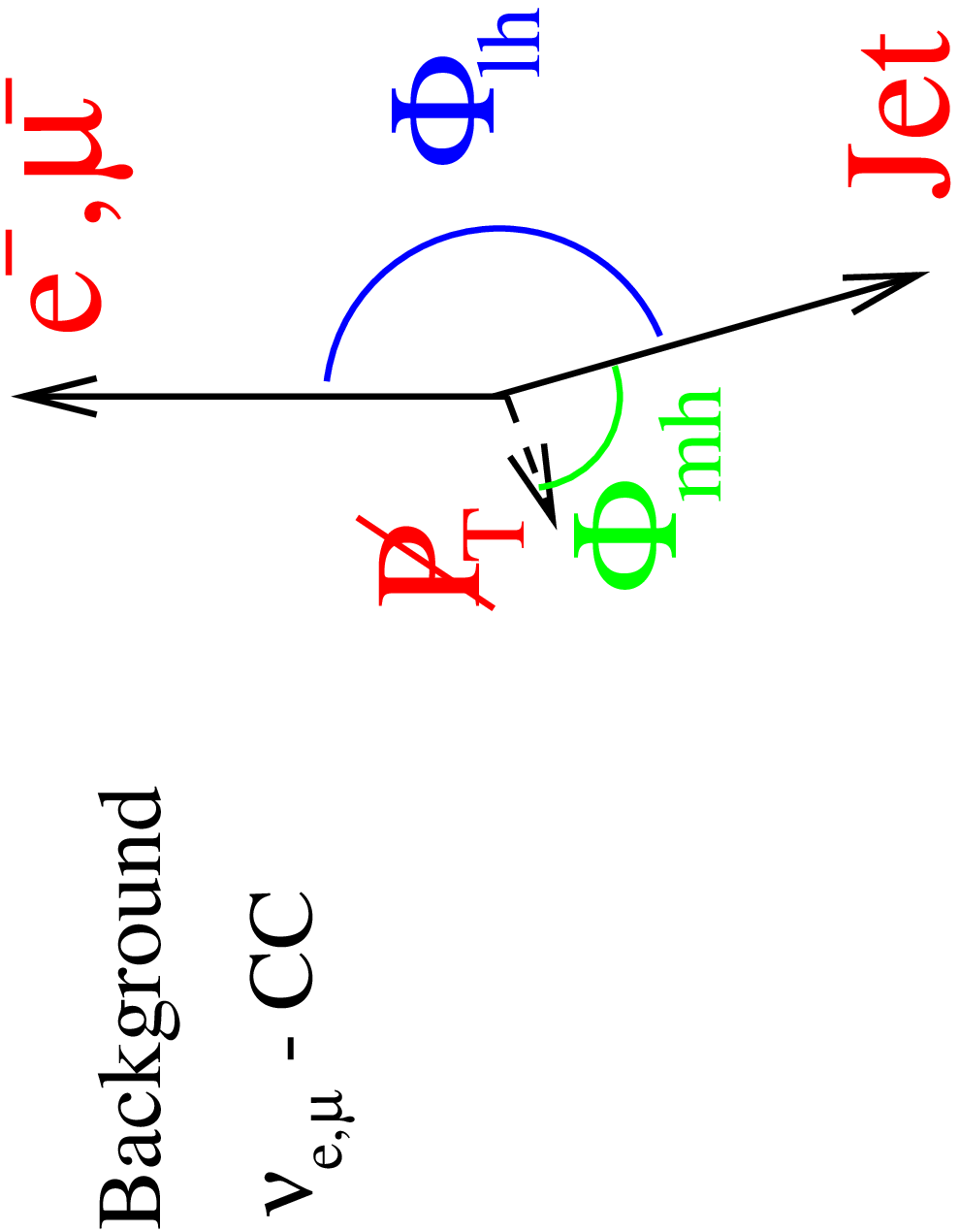,width=7cm,height=7cm}
\end{turn}} &
\mbox{\begin{turn}{270}
\epsfig{file=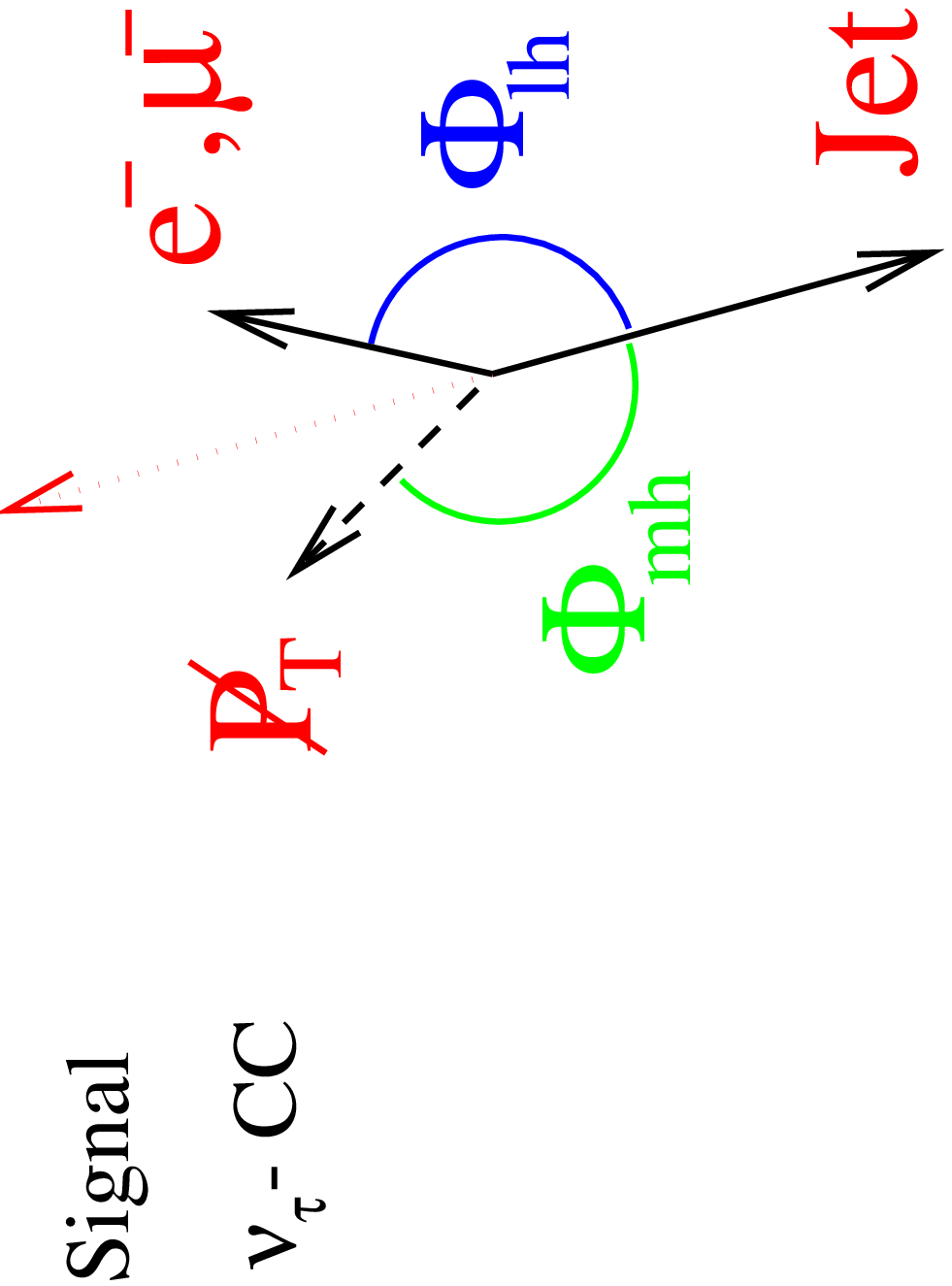,width=7cm,height=7cm}
\end{turn}}  \\
\end{tabular}
%\end{turn}}
\begin{tabular}{cc}
\epsfig{file=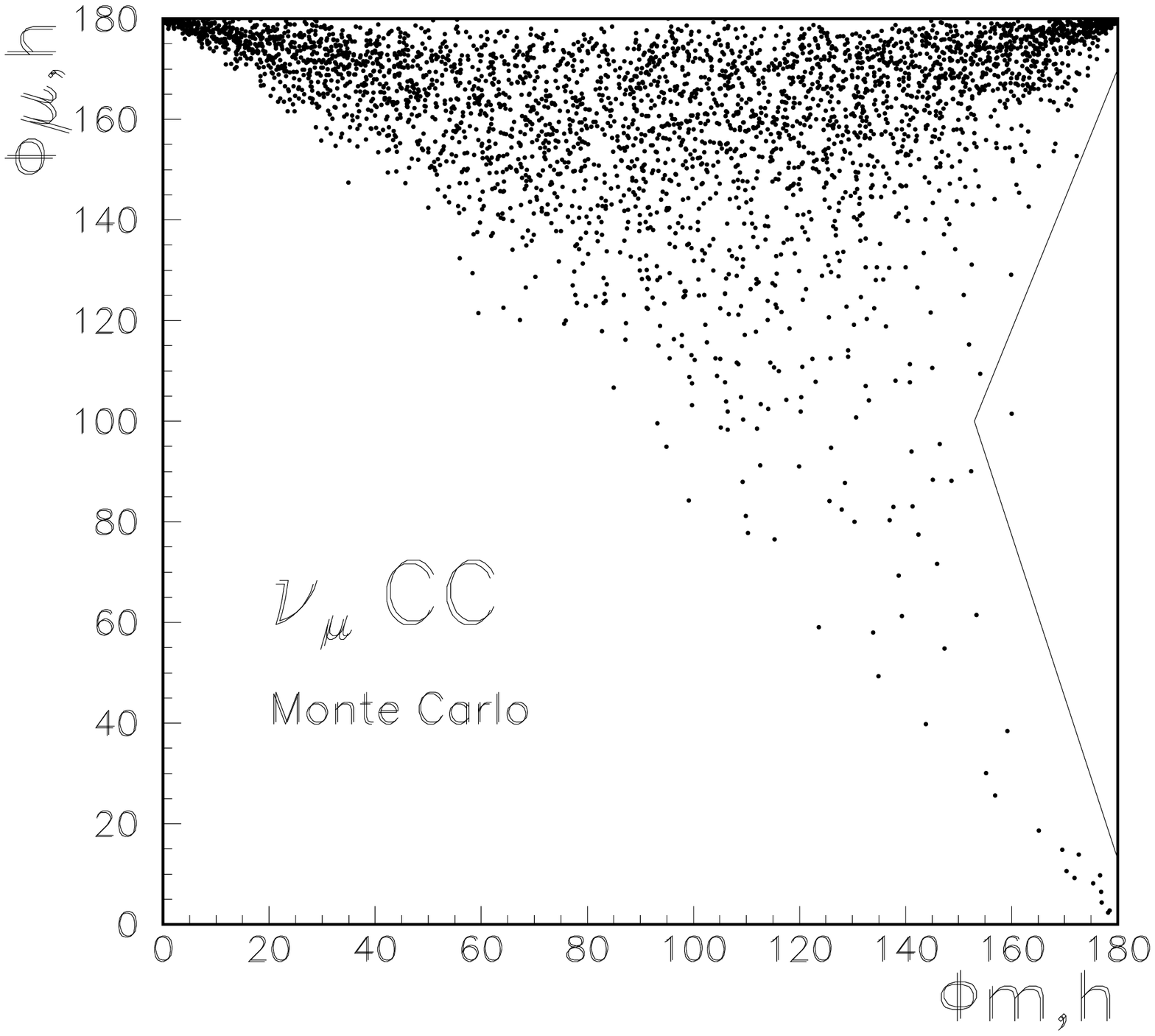,width=8cm,height=8cm} &
\epsfig{file=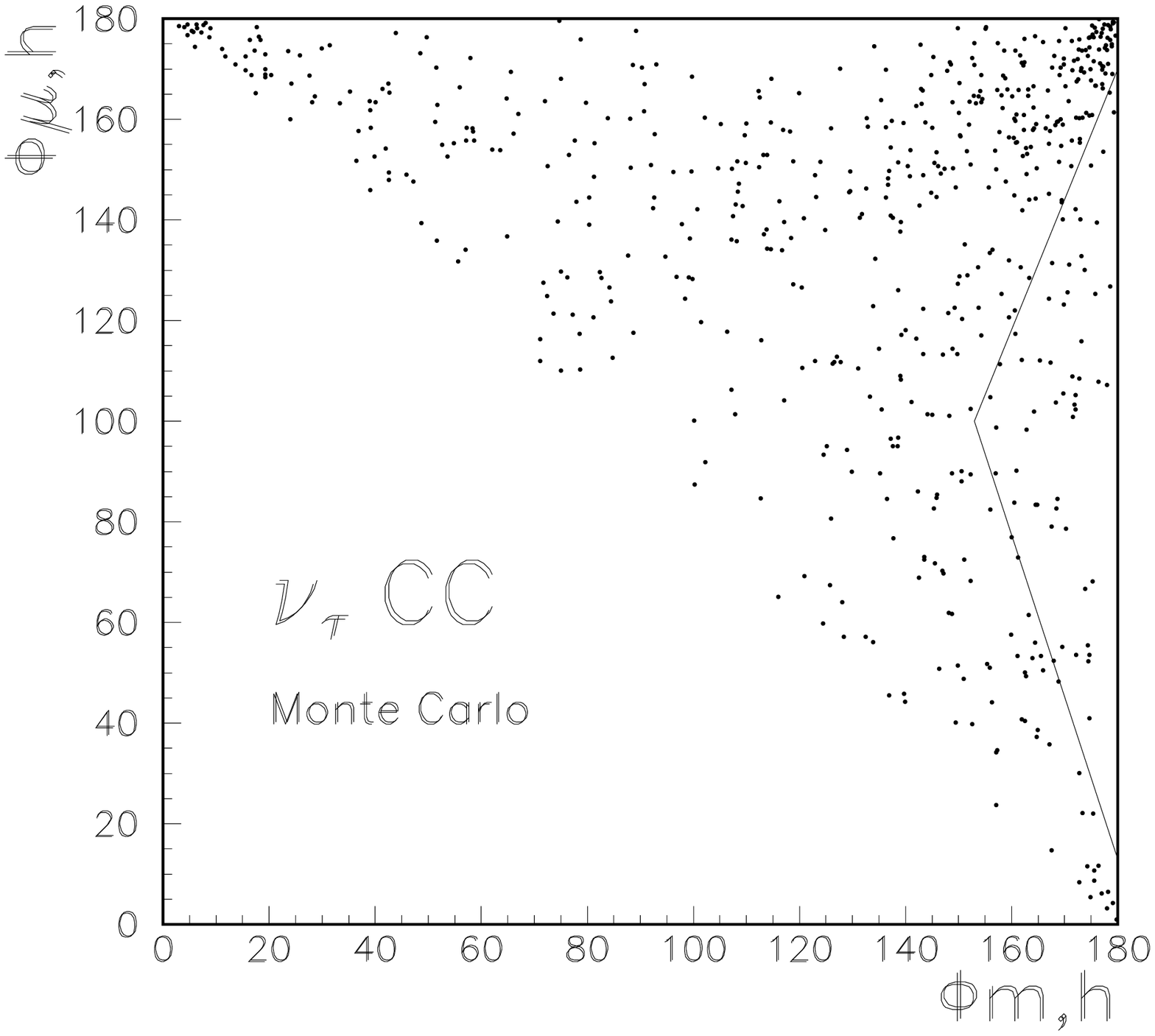,width=8cm,height=8cm}
\end{tabular}
\end{center}
\renewcommand{\baselinestretch}{1.}
\caption{\label{nomadpri} \it Definition of the kinematic variables within the NOMAD-experiment.
 For typical charged current events
the angle $\Phi_{lh}$ is close to 180 degrees (upper left). In $\tau$ - decays also the angle $\Phi_{mh}$ concentrates towards
 larger values (upper right). Double angle plots for Monte Carlo distribution for 
\nmu charged current events (lower left) and \ntau signal events (lower
right). By cutting in both angles it is possible to reduce the overwhelming charged current background
 to an acceptable level.} 
\end{figure}

High energy accelerators provide \nue beams with an average energy in the GeV region.
With respect to high energy experiments at present especially CHORUS and
NOMAD at CERN will provide new limits. They are running at the CERN wide band neutrino beam with an average energy
of around 25 GeV,
produced by 450 GeV protons accelerated in the SPS and then hitting a beryllium beam dump.
Both experiments are 823 m (CHORUS) and 835 m (NOMAD) away from the beam dump and designed to
improve the existing limits on \nmu - \ntau \oszs by an order of magnitude. 
The beam contamination of prompt \ntau from $D_s^{\pm}$-decays is of the order 10$^{-6}$.
Both experiments differ in their detection technique. While CHORUS relies on 
seeing the track of the $\tau$ - lepton and the associated decay vertex with the associated
kink because of the decay,
NOMAD relies on kinematical criteria.
\newline
 The CHORUS experiment \cite{chorusdet} (Fig. \ref{chorus}) uses
emulsions with a total mass of 800 kg segmented into 4 stacks, 8 sectors each as a main target. To determine
the vertex within the emulsion as accurate as possible systems of thin emulsion sheets and scintillating fibre trackers
are used. Behind the tracking devices follows a hexagonal air core magnet for momentum determination of hadronic tracks, 
an electromagnetic lead-scintillating fibre calorimeter with an energy resolution of 13 \% / $\sqrt{E}$ for electrons as well
as a muon spectrometer.
\begin{figure}[hhh]
\begin{center} 
\epsfig{file=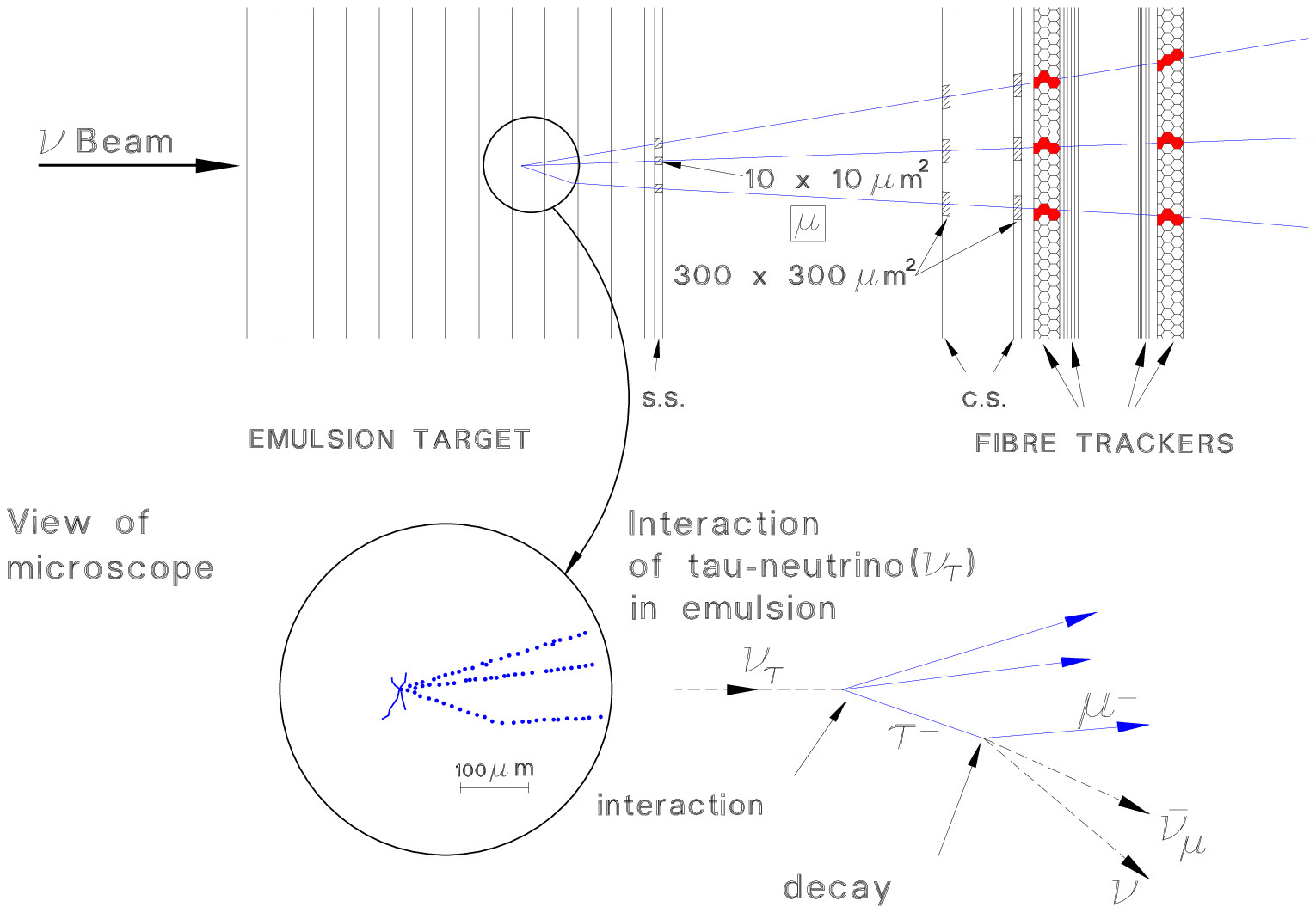,width=12cm,height=8cm}
\end{center}
\renewcommand{\baselinestretch}{1.}
\caption{\label{chorusprinz} \it Schematical view of the detection principle of CHORUS. In this case the muon
from the decay $\tau \ra \mu$ \bnmu \ntau is visible as a kink after a short $\tau$ - track.}
\end{figure}
A $\tau$ - lepton created in the emulsion by a charged current reaction is producing a track
of 
a few hundred $\mu$m. Focussing on the muonic and hadronic decay modes of the $\tau$
a signature like the one shown in Fig. \ref{chorusprinz} is expected.
The kink of the decay and the displacement of the muon from the
primary vertex should be clearly visible.
After the running period the emulsions are scanned with automatic microscopes coupled to CCDs.
The experiment has been taking data since 1994 and will continue until the end of 1997.
The present limit provided by CHORUS for the \nmu - \ntau channel for large \delm is \cite{chores}
\be
\sint < 8 \times 10^{-3} \quad (90 \% CL) 
\ee
The final goal is to reach a sensitivity down to \sint $\approx 2 \times 10^{-4}$ for large \delm .
\newline
The NOMAD experiment\cite{nomaddet} (Fig. \ref{nomad}) on the other hand relies on the
kinematics. It has as a main active target 45 drift chambers
corresponding to a total mass of 2.7 tons followed by transition radiation
and preshower detectors for e/$\pi$ separation. After an
electromagnetic calorimeter with an energy resolution of $\Delta E / E = 3.22 \% / \sqrt{E} \oplus 1.04 \%$ 
and a hadronic calorimeter five muon
chambers follow. Because most of the devices are located within a magnetic field of 0.4 T a precise momentum determination due
to the curvature of tracks is possible.
The $\tau$-lepton cannot be seen directly, the signature is determined by the decay kinematics (Fig. \ref{nomadpri}). 
The main background for the $\tau$-search are regular charged current reactions.
In normal \nmu charged current events the muon balances the hadronic final
state in transverse momentum p$_T$ with respect to the \nue beam. Hence the value for missing transverse momentum is small.
The angle $\Phi_{lh}$ between the outgoing lepton and the hadronic final state is close to 180$^{o}$ while the angle
$\Phi_{mh}$ between the missing momentum and the hadronic final state is more or less equally distributed.
In case of a $\tau$ - decay there is significant missing p$_T$ because of the escaping \ntau as well as a concentration of $\Phi_{mh}$ to larger angles because of the kinematics.
In the \nmu - \ntau channel for large \delm NOMAD gives a preliminary limit of \cite{nomadres}
% (Fig. \ref{nomadtau})
\be
\sint < 4 \times 10^{-3} \quad (90 \% CL)  \\
\ee
slightly better than the limit of E531. 
Having a good electron identification NOMAD also offers 
the possibility to search for \oszs in the \nmu - \nel channel.
A preliminary limit (Fig. \ref{nomadel}) on \nmu - \nel is available as \cite{nomadres} (for
large \delm) 
\be 
\sint < 2 \times 10^{-3} \quad (90 \% CL) 
\ee
This and a recently published CCFR result \cite{ccfr} seem to rule out the large
\delm region of the LSND evidence. 
NOMAD will continue data taking at least until end of 1997 as well. 
%\clearpage
%\begin{figure}[hhh]
%\begin{center} 
%\epsfig{file=tw3.eps,width=6cm,height=8cm}
%\end{center}
%\caption{\renewcommand{\baselinestretch}{1.} \label{nomadtau} \it \nmu - \ntau exclusion plot for \nmu - \ntau . The preliminary NOMAD limit is coming
%down at the position of the shown arrow. }
%\end{figure}
%\newpage
\clearpage
\section{Future accelerator experiments}
Possible future ideas split into two groups depending on the physical goal. One group is focussing on improving the
existing bounds by another order of magnitude with respect to CHORUS and NOMAD. This effort is motivated by the classical
\ssm which offers a \ntau in the eV-region as a good candidate for \hdm by assuming that the solar \nue problem can be solved 
by \nel - \nmu \oszsp The other group plans to increase
the source - detector distance to probe smaller \delm and to be directly comparable to atmospheric scales.
\subsection{Short and medium baseline experiments} 
Several ideas exist for a next generation of short baseline experiments. At CERN the follow up could be a
detector combining features of NOMAD and CHORUS \cite{tenor}. The idea is to use 2.4 tons of emulsion within the NOMAD magnet in
 form of 6 target modules. Each module contains an emulsion target consisting of 72 emulsion plates, as well as a set of  
large silicon microstrip detector planes and a set of honeycomb tracker planes. To verify the feasability of large silicon
detector planes maintaining excellent spatial resolution over larger areas NOMAD included a prototype (STAR) in the 1997 data taking.
For the future experiment, options to extract a \nue beam at lower energy of the proton beam (350 GeV) at the CERN SPS to reduce the prompt \ntau background are discussed as well.
At Fermilab the COSMOS (E803) experiment \cite{cosmos} is proposed to improve the sensitivity in the \nmu - \ntau channel by
one order of magnitude with respect to CHORUS and NOMAD by using the new Main Injector. It will produce a proton beam of 120 GeV resulting
in an average \nue energy of $<E_{\nu}> \approx 12$ GeV. The main target consists of two parts each containing 4 quadrants of emulsions.
Each quadrant (90 cm wide, 70 cm high and 4.9 cm thick) will contain 45 emulsion sheets summing up to a total mass of 865 kg.
The vertex finding is done with fibre trackers. The complete tracking
system is followed by a lead-glass electromagnetic calorimeter and a muon
spectrometer consisting of 6 steel slabs each followed by proportional
tubes. The distance to the beam dump will be 960 m.
 This experiment could start data taking at the beginning of the next century. Both experiments propose a sensitivity in the \nmu - \ntau channel of around 2$\times 10^{-5}$ for large \delm (\delm $>100 eV^2$). 
Also a proposal for a medium baseline search exists \cite{medium}. The CERN \nue beam used by CHORUS and NOMAD is coming up to the surface again in a distance of about 17 km
away from the beam dump. An installation of an ICARUS-type detector (liquid Ar TPC) or a 27 kton water-RICH, see
below, could be installed here.
\subsection{Long baseline experiments}
Several accelerators and underground laboratories around the world offer the possibilities to perform \lbls .
This is of special importance to probe directly the region of atmospheric \nues.
\subsubsection{KEK - \sk}
The first of these experiments will be the KEK-E362 experiment \cite{keksk} in Japan sending a \nue beam from KEK to \sk. The distance is 235 km. A 1 kt front detector, about 1 km away from the beam dump will
serve as a reference and measure the \nue spectrum. The \nue beam with
an average energy of 1 GeV is produced by a 12 GeV proton beam dump. The detection method within \sk will be identical to that of their atmospheric \nue detection.
The beamline should be finished by the end of 1998 so the experiment will start data taking in 1999.
An upgrade of KEK is planned to a 50 GeV proton beam, which could start producing data around 2004. 
\subsubsection{Fermilab - Soudan}
A big \nue program is also associated with the new Main Injector at Fermilab. The long baseline project will
send a \nue beam to the Soudan mine about 735 km away from Fermilab. Here the MINOS experiment \cite{minos} will be
installed. It also consists of a front detector located at Fermilab just upstream of COSMOS and a far detector at Soudan. The far
detector will be made of 10 kt magnetized Fe toroids in 600 layers with 4 cm thickness interrupted by about 32000
m$^2$ active detector planes in form of streamer tubes with x and y readout to get the necessary tracking informations. The project could start
at the beginning of next century.
\subsubsection{CERN - Gran Sasso}
A further program considered in Europe are \lbls using a \nue beam from CERN down to Gran Sasso Laboratory.
The distance is 732 km. Several experiments have been proposed for 
the \osz search. The first proposal is the ICARUS
experiment\cite{icarus} which will be installed in Gran Sasso anyway for the search of proton decay and 
solar neutrinos.
This liquid Ar TPC can also be used for long baseline searches. A prototyp  of 600 t is approved for installation.
A second proposal, the NOE experiment \cite{noe}, plans
to build a giant lead - scintillating fibre detector with a total mass of 4 kt. It will consist of 4 moduls
each 8m $\times$ 7.9~m $\times$ 7.3 m followed by a module for muon
identification. A third proposal is the building
of a 27 kt water-RICH detector \cite{tom}, which could be installed either inside or outside the Gran Sasso tunnel. The readout is done by 3600 HPDs with
a diameter of 250 mm and having single photon sensitivity. By using the RICH technique more complex event topologies like in standard water detectors can be investigated. 
Finally there exists a proposal for a 1kt iron-emulsion sandwich detector \cite{niwa} which could be installed either at the Fermilab-Soudan or
the CERN-Gran Sasso project. It could consist of 2240 modules arranged in 140 planes each containing 4 $\times$ 4 modules. The thickness of the iron and emulsion planes is about 1 mm.
\section{Summary and Conclusion}
\begin{figure}[hhh]
\begin{center}
\epsfig{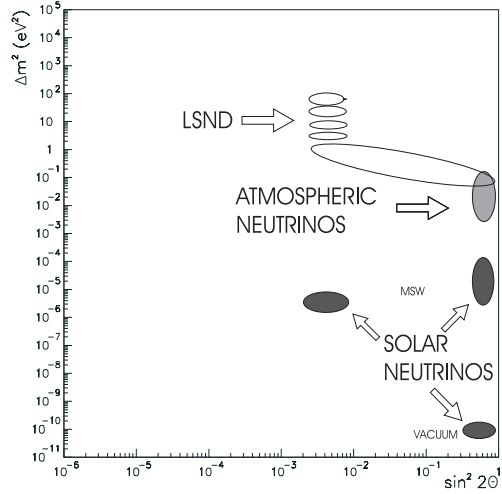}
\end{center}
\renewcommand{\baselinestretch}{1.}
\caption{\label{sum} \it Schematic presentation of parameter regions to
explain the solar (dark), atmospheric (grey) and LSND (white) results with 
\noszp}
\end{figure}
Massive \nues offer a wide range of new phenomena in \nue physics especially that
of \noszp Evidence for such \oszs comes from solar \nues, atmospheric \nues and the
\lsnd experiment. The allowed regions in the \delm - \sint parameter space are schematically shown in Fig.\ref{sum}.
Different scenarios which could be deduced from the observed evidence as well as the prefered \hdm candidate and
the observability in \sbl and \lbl are summarized in Tab.2.
\begin{figure}[hhh]
\begin{center}
\begin{tabular}{cc}
\epsfig{file=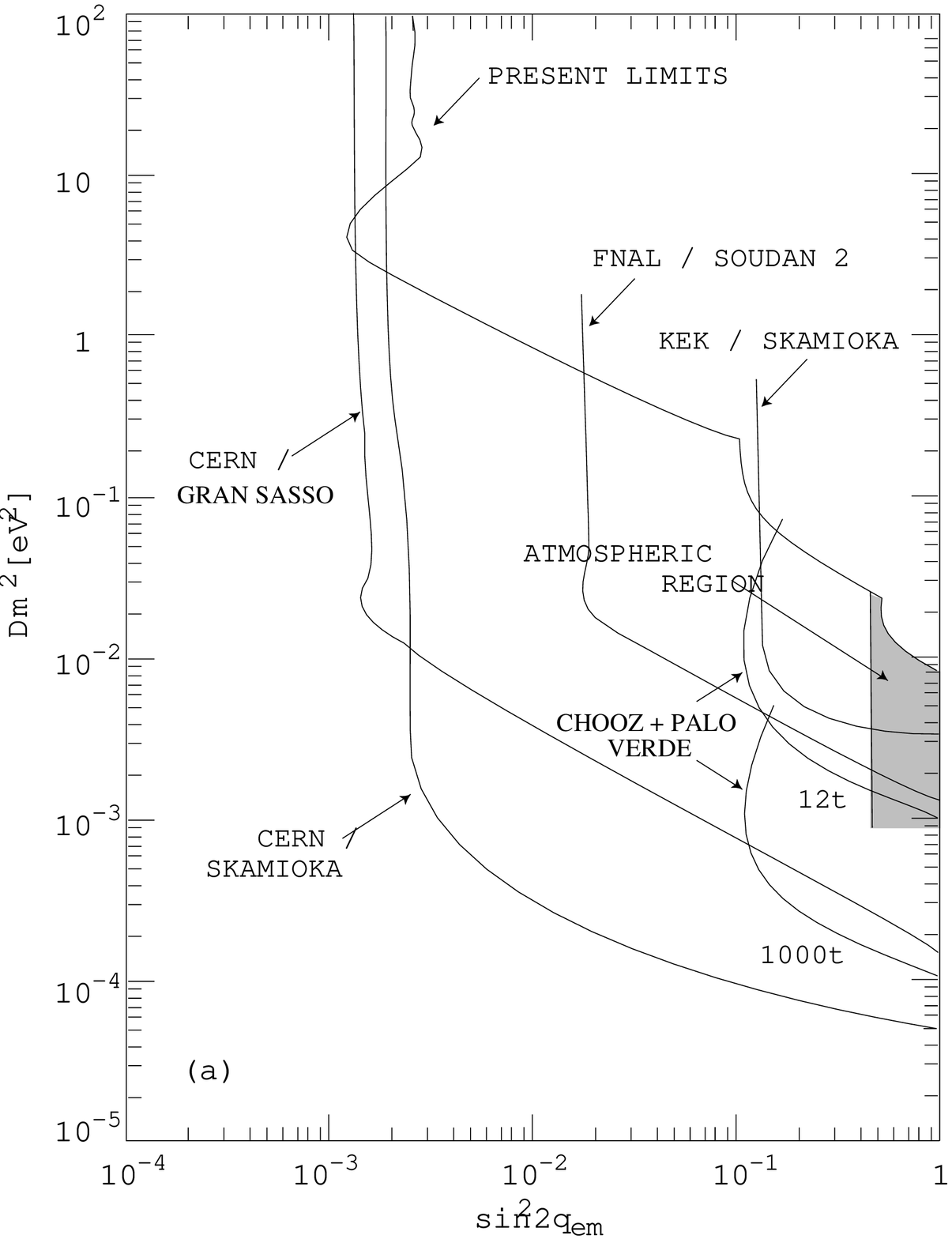,width=7cm,height=7cm}
&
\epsfig{file=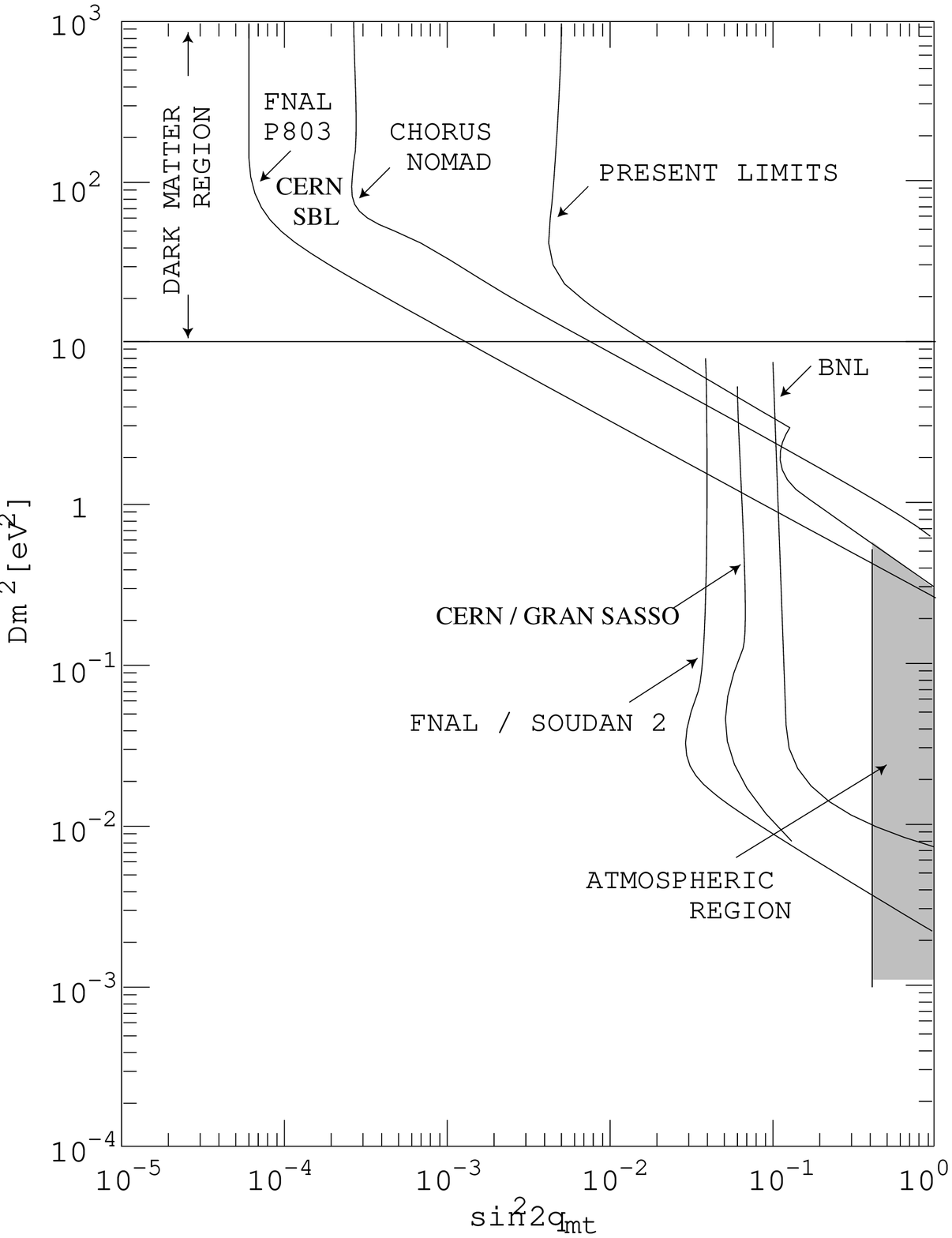,width=7cm,height=7cm}
\end{tabular}
\end{center}
\renewcommand{\baselinestretch}{1.}
\caption{\label{femu} \it Proposed exclusion regions from future reactor and accelerator experiments for the \nel - \nmu channel (left)
and the \nmu - \ntau channel (right). }
\end{figure}
\clearpage
Apart from that, terrestrial \nue experiments in form of nuclear reactors and 
high energy accelerators also exclude large parts of the parameter space because of non-observation of \osz effects. Because 
the region of the MSW-solution for solar \nues are out of range for terrestrial experiments, present
and to a large extend future \osz experiments focus on the atmospheric region and 
a proof of the \lsnd results (Fig.\ref{femu}). 
\begin{center}
{\large \begin{tabular}{|c|c|c|c|c|}
\hline solar & \multicolumn{4}{c|}{no dramatic new results} \\
\hline
atmos. & C & C & NC & NC \\
\hline 
LSND & C & NC & C & NC \\
\hline
$\mid \Delta m^2_{12} \mid $ & $\approx 1$  & $\approx 10^{-5}$ & $\approx
1$&   
 $\approx 10^{-5}$ \\
\hline
$\mid \Delta m^2_{23} \mid $ & $\approx 10^{-2}$& $\approx 10^{-2}$  &
 $\approx 1$& ? \\
\hline
$\mid \Delta m^2_{13} \mid $ & $\approx 1$& $\approx 10^{-2}$  &
 $\approx 10^{-5}$ & ? \\
\hline
\hline
Short B.L. & $\nu_e - \nu_{\tau}$ & NO & YES
& YES\\
\hline
Long B.L. & $\nu_\mu - \nu_{\tau}$ & YES & NO &
 YES \\
\hline
\hline
HDM & $ \nu_{\mu,\tau}$ & $\nu_{e,\mu,\tau}$ & $ \nu_\mu $ &
$\nu_{\tau} (\nu_{e,\mu})$ \\
\hline
\end{tabular}}
\newline
\end{center}
\medskip
{\renewcommand{\baselinestretch}{1.} \it Tab. 2: Comparison of different scenarios for the very near future depending on the
confirmation (C) or non-confirmation (NC) of the atmospheric neutrino data and
LSND results (partly compiled by L. diLella).
Also shown is the resulting observability in short and long-baseline experiments.
Neutrinos in the eV-range acting
as \hdm are shown in the last row. Only the last column allows the classical
see-saw mechanism. The first column normally cannot be explained in the framework of three \nues . Note that because of unitarity
the following relation is valid: $\Delta m^2_{12}
+ \Delta m^2_{23} = \Delta m^2_{13}$.}

\end{document}